\documentclass[11pt]{article}
\pdfoutput=1 

\usepackage{jheppub} 

\usepackage[T1]{fontenc} 

\usepackage{float} 
\usepackage{subcaption}

\title{\boldmath Janus correlators and  Heun's equation}

\author{Michael Gutperle and}
\author{Christina Yeo}

\affiliation{
Mani L. Bhaumik Institute for Theoretical Physics, Department of Physics and Astronomy,\\ University of California, Los Angeles, CA 90095, USA
}

\emailAdd{gutperle@ucla.edu}
\emailAdd{cyeo99@g.ucla.edu}

\abstract{In this paper we calculate two-point correlation functions of a massive probe scalar field in the background of the three-dimensional Janus solution. We relate the equation of motion of the scalar to  Heun's equation and use the recently obtained expressions for the connection coefficients to obtain the two-point function of operators dual to the probe scalar. From the correlators we obtain the corrections to the spectrum of boundary operators and bulk-boundary operator product coefficients to second order in the Janus deformation parameter. }

\begin{document} 
\maketitle
\flushbottom

\section{Introduction}

Heun's equation \cite{heun1888,heun-book} is a generalization of the hypergeometric equation and the most general second order 
ordinary linear differential equation with four regular singularities.  Heun's equation and its confluent limits have many applications in physics and mathematics (see e.g. 
\cite{Hortacsu:2011rr,2015arXiv151204025F}). Some important occurrences of  Heun's equation and its connection formula  are   $N=2$ supersymmetric gauge theories, namely  
the AGT correspondence \cite{Alday:2009aq}, surface operators \cite{Alday:2009fs,Awata:2010bz}, the 
Nekrasov formula for instanton counting \cite{Nekrasov:2003rj,Gaiotto:2009ma}, as well as the calculation of black hole 
perturbations, thermal correlators and quasi-normal modes in AdS/CFT \cite{Aminov:2020yma,Bonelli:2021uvf,Noda:2022zgk,Hartman:2013mia,Fitzpatrick:2014vua,Bhatta:2022wga,He:2023wcs,jia2024,BarraganAmado:2023apy,Dodelson:2022yvn,Bautista:2023sdf,Arnaudo:2024sen,Couch:2025ytb}.
Zamoldchikov \cite{Zamolodchikov86} realized that the accessory parameter of  Heun's equation can be related to  semiclassical conformal blocks \cite{Litvinov:2013sxa}. Recently \cite{Bonelli:2022ten} this relation together with the crossing symmetry of degenerate Liouville correlation functions have been used to derive connection formulas for Heun's equation and its confluent limits,  called the Trieste formula in the literature.  In \cite{Lisovyy:2022flm}  the fusion of degenerate operators  in the semiclassical limit of the Liouville CFTs has been used to derive a perturbative solution for the connection formula, and it was shown that it agrees with the Trieste formula. See \cite{daCunha:2022ewy,Desiraju:2024fmo,Iossa:2023gvh} for additional discussion of  the Trieste formula.

In this paper we present a new occurrence  of 
Heun's equation and an application of its connection formula, namely the calculation of correlation functions of a massive probe scalar in a three-dimensional Janus solution  \cite{Bak:2003jk}. The Janus solution is a simple holographic realization of a conformal interface between two conformal field theories with different  marginal deformations turned on the two sides of the interface.

The structure of this paper is as follows: In Section \ref{sec:2} we review some aspects of the Janus solution of three-dimensional AdS gravity coupled to a massless scalar field.  In Section \ref{sec:3} we consider the equation of motion of a massive probe scalar in the Janus background and derive the formula for the two-point function of the dual CFT operator using the bulk-to-boundary propagator. In Section \ref{sec:4} we show that the equation of motion for the probe scalar takes the form of  Heun's equation and determine its parameters in terms of the Janus deformation parameter and the conformal dimension of the operator dual to the probe scalar.  We also review the Trieste  formula for the connection coefficients of Heun's equation, which will be essential in calculating the bulk-boundary correlator. In Section \ref{sec:5} we present the main results of our paper.  We use the results of the previous section to calculate the two-point function in the presence of the Janus defect. As a check we consider the case of the pure $AdS_3$ solution, i.e. a trivial defect corresponding to  a vanishing Janus deformation parameter $c_J$. For nonzero $c_J$ we obtain the two-point function in an integral form in terms of the connection coefficients of Heun's equation. We use the perturbative solution for the connection coefficients to obtain the spectrum of defect operators and the bulk-boundary operator product expansion (BOPE) coefficients to lowest nontrivial order in the Janus deformation parameter. We close the paper with some discussion on possible generalizations and directions for future research. In various appendices we review some background material for the benefit of the reader, such as the basis of eigenfunctions of the Laplacian on $AdS_2$, Heun functions, and Virasoro conformal blocks. We also present details of the calculation of the bulk-boundary propagator using the Wronskian method in Appendix \ref{appendix:b}.

\section{Janus solution}
\label{sec:2}

The Janus solution \cite{Bak:2003jk} is an example of a supergravity solution describing the  holographic dual  of an interface CFT. In its simplest realization it is constructed from Einstein gravity with a negative cosmological constant coupled to a massless scalar field.   There have been many generalizations, among them the constructions of Janus solutions in various gauged supergravity theories  (see e.g. \cite{Clark:2005te,Bobev:2013yra,Suh:2011xc, Karndumri:2016tpf,Chen:2020efh,Chiodaroli:2009yw,Chiodaroli:2010ur,Gutperle:2017nwo}) and half-BPS Janus solutions in type IIB and M-theory  \cite{DHoker:2007hhe,DHoker:2007zhm,DHoker:2009lky}. Janus solutions have been important tools to study conformal defects, interfaces and their properties using holography (see e.g.  \cite{Clark:2004sb,DHoker:2006qeo,Gaiotto:2008sd}).

In this paper we will exclusively consider the simplest three-dimensional theory with a single massless scalar field \cite{Bak:2007jm} defined by the (euclidean) action
\begin{align}\label{action}
    S= {1\over 16 \pi G_3} \int d^3x \sqrt{g} \Big(R+2 -{1\over 2} \partial_\mu \phi \partial^\mu\phi\Big)\;,
\end{align}
where $G_3$ is the three-dimensional Newton's constant. 
 The equations of motion following from (\ref{action}) are given by
\begin{align}\label{eqofm}
R_{\mu\nu}+2g_{\mu\nu} -{1\over 2} \partial_\mu \phi \partial_\nu\phi=0, \quad \quad \partial_\mu(g^{\mu\nu} \sqrt{g} \partial_\nu \phi)=0\; .
\end{align}
 A Janus solution is obtained by writing the metric using  a Euclidean $AdS_2$ (i.e. $H_2$) slicing  and making the scalar only dependent on the slicing coordinate.
It is easy to verify that the following choice of metric and scalar satisfies the equations of motion
\begin{align}\label{solution}
    ds^2 ={dy^2\over 4 y^2(1-y)^2}+ f(y) {dz^2+d\tau^2 \over z^2}, \quad \quad {d\phi\over dy}= {2 c_J \over y(1-y) } {1\over f(y)}\;.
\end{align}
If one chooses $f$ as follows
\begin{align}
     f(y) = {1-c_J +2 c_J y \over 4 y(1-y)}\;,
\end{align}
the coordinate $y$ takes values  in the unit interval $y\in[0,1]$.  By integrating the second equation of (\ref{solution}) one obtains
\begin{align}
    \phi(y)= \phi_0 + \ln\big(1- c_J+ 2 c_J y\big)\;.
\end{align}
The holographic interpretation of the solution is obtained by considering the asymptotic boundary components of the metric (\ref{solution}).  There are three boundary components in an $H_2$ sliced Janus geometry. Two boundary components are given by $y\to 0$ and $y\to 1$. They correspond to two half spaces which are glued together at an interface, which can be identified with the third boundary component of $H_2$ at $z\to 0$ \cite{Bak:2003jk,Freedman:2003ax,Clark:2004sb}. The scalar field takes different values on the two sides depending on  the integration constant $c_J$, and the asymptotic value of the scalar field jumps by
\begin{align}\label{scalarjump}
    \Delta \phi = \phi|_{y=1}-\phi|_{y=0} = \ln\left({ 1+c_J\over 1-c_J}\right)\;.
\end{align}
In holography a massless scalar is dual to  a marginal operator on the CFT side, hence the  interpretation in the Janus solution is given by  a two-dimensional CFT whose modulus takes different values across the interface. A useful toy model is given by considering a CFT given by a compact scalar where the compactification radius $R$ jumps across the defect \cite{Bachas:2001vj}. Note that if we select $c_J=0$, the scalar is constant and hence  the jump vanishes, yielding the $AdS_3$ vacuum with unit curvature.
It follows from (\ref{scalarjump}) that there are  extremal values of $c_J=\pm 1$, where the jump diverges and the spacetime becomes singular \cite{Gutperle:2012hy}.

\section{Probe scalar in Janus background}
\label{sec:3}

In a bottom-up approach to AdS/CFT, we can include additional  operators in the CFT by adding fields to the three-dimensional action.  
The probe approximation is characterized by considering the free action for such fields and neglecting higher order couplings as well as  gravitational backreaction. This approximation is sufficient for calculating two-point functions of the dual operators.

\subsection{Equation of motion for the probe scalar}

In particular, scalar primaries of scaling dimension $\Delta$ are described by adding  a massive scalar field
\begin{align}\label{action1}
    S_{probe}=  -\int d^3 x \sqrt{g} \Big( {1\over 2}  \partial_\mu \psi \partial^\mu \psi + {1\over 2} m^2 \psi^2 \Big)\;,
\end{align}
where the scaling dimension $\Delta=\Delta_+$  is expressed in terms of the mass $m$ of the scalar field
 by
 \begin{align}\label{delpm}
\Delta_\pm (\Delta_\pm -2) = m^2, \quad \quad \Delta_{\pm} =1\pm \sqrt{1+m^2}\;.
\end{align}
Note that the mass squared can be negative, but is bounded by the Breitenlohner-Freedman  \cite{Breitenlohner:1982jf} bound $m^2>-1$ for an $AdS_3$ space with unit curvature radius. These relations are for the standard quantization, which is what we adopt in this paper. The alternate quantization of fields with $-1<m^2<0$ identifies the scalar to a double trace operator \cite{Klebanov:1999tb}; see \cite{Melby-Thompson:2017aip} for a discussion of interfaces and double trace operators. In the following we will consider the massive scalar field (\ref{action1}) in the probe approximation, i.e. we will neglect the backreaction of the scalar on the metric as well as its coupling to other fields and nonlinear terms in the scalar action.  This allows us to calculate the two-point function of the operators dual to the massive scalar fields in the Janus background exactly. One has to go beyond the probe approximation to calculate higher point functions or correlation functions, but this will be beyond the scope of this paper.

We can expand the scalar field in the basis of eigenfunctions $g_{s,k} (z,t)$ of the Laplacian of $H_2$
\begin{align}\label{ads2exp}
    \psi(y,z,\tau ) &= \int dk \int ds\;  \mu(s) h_{s}(y) g_{s,k} (z,\tau ) \;, 
\end{align}
with $s,k$ parameterizing the $H_2$ Laplacian eigenvalue and its momentum along the $\tau$ direction,
\begin{align}
    z^2\big( \partial_z^2+ \partial_\tau^2\big)g_{s,k} (z,\tau)= -(s^2+{1\over 4})  g_{s,k} (z,\tau )\;.
\end{align}
See Appendix \ref{appendix:a} for a review of this basis, including the explicit form of the integration measure, eigenfunctions, and completeness/orthogonality relations.
With this expansion, the equation of motion for the probe scalar (\ref{action1}) in the Janus background becomes an ordinary differential equation for $h_{s}(y)$.

\begin{align} \label{EOM_h}
    h_s''(y)  +{2c_J\over 1-c_J+ 2 c_J y} h_s'(y)- {s^2+{1\over 4}\over y(1-y)(1-c_J + 2 c_J y)} h_s(y) + {(2-\Delta)\Delta \over 4 y^2(1-y)^2 }h_s(y)&=0\;.
\end{align}
As we show below, this equation is  Heun's equation, i.e. a second  order ODE of Fuchsian type with four regular singularities located at $y=0,1, \infty,$ and $y= {c_J-1\over 2c_J}$.  The singular points $y=0,1$ correspond to the two asymptotic $AdS$ boundaries.

The behavior  of a solution $h_s$ near the asymptotic AdS boundary components  is determined by the scaling dimension and can be expressed as  power series expansion
\begin{align}\label{boundex}
 y\to 0:&\quad     h_s \sim  a_{+,s} y ^{{1\over 2}\Delta_+} + a_{-,s} y ^{{1\over 2}\Delta_-}+ \cdots  \;, \nonumber\\
 y\to 1:&\quad     h_s \sim  b_{+,s} (1-y) ^{{1\over 2}\Delta_+} + b_{-,s} (1-y) ^{{1\over 2}\Delta_-}+ \cdots \;,
\end{align}
where the exponents $\Delta_\pm$ are given by (\ref{delpm}).  
If $\Delta_+ -\Delta_-$ is an integer, the two terms in the expansion are not independent and one has to add a logarithmic term. We will comment on this case in the discussion, but in the following we  will assume that the expansion takes the form (\ref{boundex}). 

The most general solution of (\ref{EOM_h}) is then determined by choosing  $a_{\pm,s}$ or $b_{\pm,s}$, then obtaining the other coefficient by the connection formula for (\ref{EOM_h}). In the following sections we will discuss this connection formula and use it to calculate the two-point function for our probe scalar.

\subsection{Two point correlation functions}
The AdS/CFT correspondence identifies the generating function of correlation functions with the gravitational partition function with the corresponding boundary sources for the bulk fields turned on.  
The finite on-shell action in AdS/CFT is obtained using the machinery of holographic renormalization \cite{Bianchi:2001kw, Kraus:1999di, Skenderis:2002wp}.  For the probe scalar action, this entails integrating (\ref{action1}) by parts, regularizing the action by introducing a cutoff  of the AdS boundaries at $y=\epsilon$ and $y=1-\epsilon$, adding a  counterterm to make the action finite\footnote{This counterterm is sufficient for $\Delta < 3$ \cite{Skenderis:2002wp}, in general additional counterterms can be added to make the on-shell action finite.}, and then taking the cutoff to zero. This gives
\begin{align} \label{Srenorm}
    S_\mathrm{ren} &= \lim_{\epsilon\to 0}\int dz d\tau   \left. \Big( -{1\over 2}\sqrt{g} g^{yy} \psi \partial_y \psi -{1\over 2} \sqrt{\gamma} \Delta_- \psi^2\Big)\right|_{y=\epsilon}^{y=1-\epsilon}\;.
\end{align}
Here $\gamma_{ab}$ is the induced metric of the $H_2$ factor.

For a Janus geometry, we can turn on sources on either side of the boundary corresponding to the two half spaces bisected by the interface. In the following, we denote the $y=0$ boundary as the L(eft)  and the $y=1$ boundary as the R(right).  The bulk field can be expressed in terms of the sources
\begin{align} \label{bulkfield}
    \psi(y,z,t) = \int {dz' \; d\tau'\over z'^2} \Big( K_L(y,z,\tau| z',\tau') J_L(z',\tau')+ K_R(y,z,\tau| z',\tau') J_R(z',\tau')\Big)\;,
\end{align}
where  $K_{L/R}$ corresponds to the bulk-to-boundary propagator (\ref{bulkbndyprop}) with the source on the left/right half-space, respectively .

The two-point correlation functions  of the operators dual to the probe scalar are then given by derivatives of the on-shell action with respect to the sources $J_{L/R}$
\begin{align} \label{2pt}
    \langle O(z,\tau) O(z', \tau')\rangle_{RR} &= {\delta^2 S_{ren} \over \delta J_R(z,\tau) J_R(z',\tau') }\;, \nonumber \\
      \langle O(z,\tau) O(z', \tau')\rangle_{LL} &= {\delta^2 S_{ren} \over \delta J_L(z,\tau) J_L(z',\tau') }\;, \nonumber\\
        \langle O(z,\tau) O(z', \tau')\rangle_{LR} &= {\delta^2 S_{ren} \over \delta J_L(z,\tau) J_R(z',\tau') }\;.
\end{align}
The coordinates $z,\tau$ parameterize a half space and the $RR$ and $LL$ correlators correspond to both operators inserted on the same half space, while the $LR$ correlator corresponds to one operator inserted on the left half space and one on the right.

\subsection{Bulk boundary operator product expansion}\label{sec:3.3}

The two-point functions (\ref{2pt}) in the presence of a co-dimension one defect allow for bulk and boundary operator expansions, which have been discussed for CFTs in general dimensions in \cite{McAvity:1995zd,Billo:2016cpy,Liendo:2012hy}. We are specializing to the $d=2$ case which is relevant for the $AdS_3/CFT_2$ context. The correlation function of two bulk operators can be expressed in terms of the cross-ratio

\begin{align}
\langle O_1(z,\tau) O_2(z',\tau')\rangle =  {1\over |z|^{\Delta_1} |z'|^{\Delta_2}}  f{(\xi}) \;. \label{2ptitocr}
\end{align}
For an interface/defect  localized at $z=0$, the distinction of the $LL,LR,RR$ correlation functions, i.e. whether the bulk operators are inserted on the same or opposite sides of the defect can be encoded in the cross ratio
\begin{align}\label{allxi}
    \xi_{LL,RR}&= {(\tau-\tau')^2  +(z-z')^2 \over 4z z'}, \quad \quad    \xi_{LR,RL}= {(\tau-\tau')^2  +(z+z')^2 \over 4zz'}\;.
\end{align}
Operators inserted on the same  side can be brought close together away from the defect, which corresponds to the standard bulk OPE and the limit $\xi_{LL,RR} \to 0$.  Note that such a limit is impossible for operators inserted on opposite sides of the defect.
We are interested in the limit of the bulk operators close to the defect/interface, which corresponds to the limit $\xi\to \infty$ for all cross ratios (\ref{allxi}), so we drop the subscript in the following. This lets us express (\ref{2ptitocr}) in terms of the bulk-boundary conformal block \cite{McAvity:1995zd,Billo:2016cpy}
\begin{align} \label{bope}
    f(\xi)&\sim \sum_{\hat O_i} b_{O_1 \hat O_i  } b_{O_2\hat O_i }  \;  {1\over |\xi|^{\hat \Delta_{i}}} {}_2 F_1(\hat \Delta_{i},\hat \Delta_{i},2\hat \Delta_{i},-{1\over \xi})\;,
\end{align}
where $\hat \Delta_i$ denote the conformal dimensions of the boundary operators appearing in the bulk-boundary operator product expansion (BOPE) and $ b_{O \hat O_i  }$ are the BOPE coefficients. In this paper we will use holography to calculate the spectrum of boundary operators $\hat \Delta_i$ as well as the BOPE coefficients for a probe scalar field in the Janus interface theory.

\section{Heun connection formula}
\label{sec:4}

In this section we relate the equation of motion for the probe scalar in the Janus background to  Heun's equation, and present the relevant connection formulas which will be used to calculate the two-point correlators holographically.  Heun's equation is a natural generalization of the hypergeometric differential equation in the sense that it is a second order ordinary differential equation of Fuchsian type with four instead of three regular singular points. 
In particular we are interested in the connection formula which relates Frobenius expansions around different singular points. This formula was obtained in \cite{Bonelli:2022ten, Lisovyy:2022flm,Consoli:2022eey,jia2024} from the  fusion transformation  of degenerate Virasoro blocks in the semiclassical limit.

\subsection{Heun's equation}
After rescaling $h_s$  by a factor
\begin{align} \label{solnresc}
    h_s(y) = (1-c_J+2 c_J y)^{-{1\over 2}} \rho_s(y) ,
\end{align}
 the probe scalar equation (\ref{EOM_h}) becomes  Heun's equation in its normal form
\begin{align} \label{probeHeun}
    \rho_s''(y)+ T(y)\rho_s(y) =0
\end{align}
with
\begin{align}\label{heunoper}
    T(y) = {\delta_0\over y^2} + {\delta_1 \over (y-1)^2}+ {\delta_t \over \big( y- t\big)^2} + {\delta_\infty-\delta_0-\delta_1-\delta_t\over y(y-1)}+{(t-1)\chi \over y(y-1)(y- t)}\;.
\end{align}
The parameters can be expressed in terms of the scaling dimension of the probe scalar, the Janus deformation parameter, and the eigenvalue of the $H_2$ Laplacian
\begin{align}\label{paraheun}
    t=  {-1+c_J \over 2c_J}, \quad \delta_0=\delta_1={\Delta \over 2}(1-{\Delta \over 2}), \quad \delta_t=\delta_\infty={1\over 4}, \quad \chi={-s^2 -{1\over 4} \over 1+c_J}\;.
\end{align}
Note that the $H_2$ Laplacian   eigenvalue $-s^2-{1\over 4}$ only enters through the accessory parameter $\chi$, whose appearance  is a new feature for  Heun's equation compared to the hypergoemetric differential equation.

The parameters $\delta_i=({1\over 2}+\theta_i)({1\over 2}-\theta_i), i=0,1,t,\infty$ have a nice interpretation. ${1\over 2}\pm \theta_i$ are the critical exponents at each singularity, and the parameters $\theta_i$ will be related to the Liouville momenta used in the fusion expansion, which is reviewed in Appendix \ref{appendix:d}. It follows from (\ref{paraheun}) that the critical exponents at $y=0,1$, which correspond to the AdS boundaries, depend on the scaling dimension $\Delta$. There is an overall sign ambiguity when $\theta_i \neq 0$ and a choice between $\Delta=\Delta_\pm$; for later notational ease, we'll choose $\theta_{0,1}={\Delta_- - 1\over 2}$.

Solutions to (\ref{probeHeun}) near $y=0$ can be expanded in terms of two linearly independent functions, the so-called Heun functions
\begin{align}
    \rho_{s,+}^{L}&= y^{\Delta_+\over2} H\ell(t,\chi(1-t)+\Delta_-({1\over 2} +t(\Delta_--1));\Delta_-,\Delta_-,\Delta_-,\Delta_-;y) \;,\\
    \rho_{s,-}^{L}&= y^{\Delta_-\over2} H\ell\left(t,1-{\Delta_- \over 2} + (1-t)\chi;1,1,2-\Delta_-,\Delta_-;y\right) \;,
    \end{align}
    and similarly for the solutions near $y=1$
    \begin{align}
    \rho_{s,+}^{R}&= (1-y)^{\Delta_+\over 2} H\ell(1-t,(t-{1\over 2})\Delta_- + (1-t)(\Delta_-^2-\chi);\Delta_-,\Delta_-,\Delta_-,\Delta_-;1-y)\;,\\
    \rho_{s,-}^{R}&= (1-y)^{\Delta_-\over2} H\ell\left(1-t,1-{\Delta_-\over 2}-(1-t)\chi;1,1,2-\Delta_-,\Delta_-;1-y\right) \;,
\end{align}
with superscripts L and R chosen to indicate the left and right AdS boundaries and $\pm$ denoting the two independent solutions. The Heun function  $H \ell(a, q ; \alpha, \beta, \gamma, \delta ; z)$ is a Fuchs-Frobenius solution to Heun's equation and  the explicit form of $H \ell$ as a power series is given in Appendix \ref{appendix:c}.

\subsection{Connection formula}
In the following  we  calculate  the two-point correlation functions from the bulk-boundary  propagator. An important ingredient in the construction is the connection formula, which relates the two linearly independent normalizable and non-normalizable solutions $\rho^L_{s, \pm}$ at the boundary at $y=0$ to the normalizable and non-normalizable solutions $\rho^R_{s, \pm}$ at the other boundary at  $y=1$

\begin{equation} \label{connform}
    \rho_{s,\epsilon}^{L}(y)=\sum_{\epsilon'=\pm} \mathsf{C}_{\epsilon\epsilon'} \rho_{s,\epsilon'}^{R}(y)\;.
\end{equation}
The connection coefficients can be computed by two different methods: from semiclassical Virasoro conformal blocks (Trieste formula) \cite{Bonelli:2022ten}, or from large-order asymptotics of Frobenius expansion coefficients (Sch\"afke-Schmidt formula) \cite{Lisovyy:2022flm}. We adopt the former method for our calculation of the two-point function. We review some details of the CFT calculation, originally presented in \cite{Bonelli:2022ten,Lisovyy:2022flm,Consoli:2022eey}, in Appendix \ref{appendix:d}, and show the agreement between the two methods to first non-trivial order in Appendix \ref{appendix:e}.
In the following we present the results which are relevant for the calculation of the two-point function.

The connection formula can be obtained from the semiclassical conformal block in the s-channel \cite{jia2024}, whose ${1 \over t}$ expansion has the form
\begin{align}\label{wexp}
    \mathcal{W}(t) &= (\delta_{\infty}-\delta_{\sigma}-\delta_t) \ln(t) + \sum_{k=1}^\infty {\mathcal W}_n(\{\theta_i\},\sigma) {1\over t^k}\;.
\end{align}
We will work to order ${1\over t^2}$ in this paper. The first two terms in the expansion are given by
\begin{align}
    \mathcal{W}_1(\{\theta_i\},\sigma)&=  {(\delta_{\sigma}-\delta_0+\delta_1)(\delta_{\sigma}-\delta_{\infty}+\delta_t) \over 2 \delta_{\sigma}} \;, \nonumber\\ 
\mathcal{W}_2(\{\theta_i\},\sigma)&={(\delta_{\sigma}-\delta_0+\delta_1)^2(\delta_{\sigma}-\delta_{\infty}+\delta_t)\over 8 \delta_\sigma^2}\left(  {1\over \delta_\sigma-\delta_0 +\delta_1} + {1\over \delta_\sigma -\delta_\infty +\delta_t} - {1\over 2\delta_\sigma} \right)
    \nonumber\\
    &+ {\big(\delta_\sigma^2 + 2 \delta_\sigma(\delta_0+\delta_1)- 3 (\delta_0-\delta_1)^2\big) \big(\delta_\sigma^2 + 2 \delta_\sigma(\delta_\infty +\delta_t)- 3 (\delta_\infty-\delta_t)^2\big)  \over 16 \delta_\sigma^2(4\delta_\sigma+3)}\;.
\end{align}
Higher order terms can in principle be calculated using a recursion formula first obtained in \cite{Zamolodchikov86} and worked out up to third order in \cite{Ferrari:2012gc,Be_ken_2020,Piatek:2021aiz}.  The expansion  becomes rather unwieldy quickly and the second order is all there is needed to obtain results to first nontrivial order in the Janus perturbation.

The values of the quantities $\delta_i={1\over 4}-\theta_i^2, i=0,1,t,\infty$ are given in (\ref{paraheun}). The fifth parameter $\delta_\sigma = {1\over 4} - {\sigma} ^2$ and $\sigma$ is  related to the momentum of the internal leg of the 4-point Virasoro block.  It can be connected to the  accessory parameter $\chi$  which appears in  Heun's equation (\ref{heunoper}) via the Zamolodchikov relation
\begin{align}\label{zamrel}
    t {\partial W\over \partial t} = \chi\;.
\end{align}
This relation can be used to determine $\sigma$  as a power series in $1/t$
 \begin{align} \label{liouvintmom}
    \sigma &= \omega  + \sum_{k=1}^{\infty} [\sigma]_k t^{-k} \;,
    \end{align}
where the leading Liouville momentum is fixed by the leading OPE term in $\mathcal{W}(t)$
\begin{equation} \label{Lisovyy2.24}
    \omega^2 =\theta_{\infty} ^2 - \theta_t ^2 +\chi + {1 \over 4}\;.
\end{equation}
We use the expansions (\ref{wexp}) and (\ref{liouvintmom}) to solve (\ref{zamrel}) order by order in $1/t$.
The connection coefficients (\ref{connform}) depend on $\sigma$ and the semi-classical conformal blocks and are given by  (see Appendix \ref{appendix:d} for details)
\begin{equation} \label{trieste}
\mathsf{C}_{\epsilon\epsilon'}=\frac{\Gamma\left(1-2 \epsilon \theta_0\right)\Gamma\left(2 \epsilon' \theta_1\right)}{\Gamma\left(\frac{1}{2}-\epsilon \theta_0+\epsilon' \theta_1\pm\sigma\right)}\exp\left[\frac{1}{2}\left(\epsilon' \partial_{\theta_1}-\epsilon \partial_{\theta_0}\right)\mathcal{W}(t)\right]\;.
\end{equation}
    
Here we've introduced the shorthand $\Gamma(a\pm b)=\Gamma(a+b)\Gamma(a-b)$, and will adhere to this notation for the rest of this paper. Both the accessory parameter $\chi$ as well as $t$ depend on the Janus deformation parameter $c_J$ via the relations (\ref{paraheun}). In particular, an expansion in small $c_J$ corresponds to an expansion in $1/t$.  Here we present the expansion of $\sigma$ to second order in the Janus deformation parameter $c_J$ instead of $1/t$, which is needed in the next section

\begin{align}\label{sigmas}
    \sigma&= i\; s + i {11+ 48 s^4 +20(\Delta_--2)\Delta_- + 8 s^2 (7+2(\Delta_--2)\Delta_-)\over 256 s(s^2+1)} c_J^2+ \mathcal{O}(c_J^3)\;.
\end{align}

\section{The two-point correlator  and the defect operator spectrum}\label{sec:5}

In this section we calculate the holographic two-point function for the probe scalar (\ref{2pt}) using the bulk-boundary propagator (\ref{bulkfield}). We present the details of the construction of the propagator using the Wronskian method in Appendix \ref{appendix:b}. The main ingredients are the two linearly independent solutions which are normalizable on the right and left boundary respectively, which are normalized as follows
\begin{align} \label{normsoln}
    N_s^R &=\rho_{s,+}^R, \quad \quad   \lim_{y\to 1 } N_s^R=
     (1-y)^{\Delta_+\over 2} +\cdots\;, \nonumber\\
    N_s^L &=\rho_{s,+}^L,\quad \quad  
\lim_{y\to 0}  N_s^L =y^{\Delta_+\over 2}  +\cdots \;.
\end{align}
Using the connection formula (\ref{connform}), the expansion of the normalizable modes $N^{L,R}_s$  near the opposite boundaries takes the following form
\begin{align} \label{nnormsoln}
    N_s^R& \sim -\mathsf{C}_{--} y^{\Delta_+\over 2} +\mathsf{C_{+-}} y^{\Delta_-\over 2}  +\cdots& y\to 0 \;,\nonumber\\
    N_s^L&\sim  \mathsf{C}_{++} (1-y)^{\Delta_+\over 2} + \mathsf{C}_{+-} (1-y)^{\Delta_-\over 2}\ +\cdots & y\to 1\;.
\end{align}
Equipped with (\ref{normsoln}) and (\ref{nnormsoln}), we find the near-boundary behavior of the bulk field (\ref{bulkfield}) as a function of the sources at the right boundary boundaries to be (see Appendix \ref{appendix:b} for details)
\begin{align} \label{fieldR}
    \lim _{y \rightarrow 1} \psi (y,z,\tau)
    =& R_-(z,\tau)(1-y)^{\Delta_-/2}+R_+(z,\tau)(1-y)^{\Delta_+/2} + \cdots \;,
\end{align}
    with 
\begin{align}
    R_-(z,\tau) =& {1\over \sqrt{1+c_J}} J_R(z,\tau) \;,\\
    R_+(z, \tau) =& {1\over \sqrt{1+c_J}} \int {dz' d\tau' \over (z')^2} \int_{-\infty}^{\infty} dk \int_{0}^{\infty} ds ~ \mu (s) g^*_{k,s}(z,\tau)  g_{k,s} (z',\tau')\nonumber\\
    &\times \Big( {\mathsf{C}_{++}\over \mathsf{C}_{+-}} J_R(z',\tau')+{1\over \mathsf{C}_{+-}} J_L(z',\tau') \Big)\;,
\end{align}
and for the left boundary we have
\begin{align} \label{fieldL}
    \lim _{y \rightarrow 0} \psi (y,z,\tau)
    =& L_-(z,\tau)y^{\Delta_-/2}+L_+(z,\tau) y^{\Delta_+/2} + \cdots\;.
\end{align}
with
\begin{align}
  L_-(z,\tau) =& {1\over \sqrt{1-c_J}}  J_L(z,\tau) \;,\nonumber\\
    L_+(z,\tau) =& {1\over \sqrt{1-c_J}} \int {dz' d\tau' \over (z')^2} \int_{-\infty}^{\infty} dk \int_{0}^{\infty} ds ~ \mu (s) g^*_{k,s}(z,\tau)  g_{k,s} (z',\tau')\nonumber\\
    &\times \Big( -{\mathsf{C}_{--}\over \mathsf{C}_{+-}} J_L(z',\tau')+{1\over \mathsf{C}_{+-}} J_R(z',\tau') \Big).
\end{align}
The renormalized on-shell action (\ref{Srenorm}) is given by terms evaluated at the boundaries  $y=0$ and $y=1$ and can then be written as
\begin{align}
S_\mathrm{ren} = {1\over 4}(\Delta_+-\Delta_-) \int {dz d\tau \over z^2}\Big( (1+c_J) R_-(z,\tau)R_+(z,\tau) + (1-c_J) L_-(z,\tau)L_+(z,\tau) \Big)\;.
\end{align}
The two-point function  (\ref{2pt})  is obtained  by taking derivatives with respect to the sources $J_{L/R}$ at the respective boundaries and can be expressed as an integral over the eigenvalues of the Laplacian on $H_2$  in terms of the connection coefficients as follows:
\begin{align}
     \langle O(z,\tau) O(z', \tau')\rangle_{RR} =& {(\Delta_+-\Delta_-) \over 2} \int_{-\infty}^{\infty} dk \int_{0}^{\infty} ds ~ \mu (s)~ g^*_{k,s}(z,\tau)  g_{k,s} (z',\tau') {\mathsf{C}_{++} \over  \mathsf{C}_{+-}}\;, \nonumber \\
     \langle O(z,\tau) O(z', \tau')\rangle_{LL} =& {(\Delta_+-\Delta_-) \over 2} \int_{-\infty}^{\infty} dk \int_{0}^{\infty} ds ~ \mu (s)~ g^*_{k,s}(z,\tau)  g_{k,s} (z',\tau') {-\mathsf{C}_{--} \over  \mathsf{C}_{+-}}\;, \nonumber\\
     \langle O(z,\tau) O(z', \tau')\rangle_{RL/LR} =& {(\Delta_+-\Delta_-) \over 2} \int_{-\infty}^{\infty} dk \int_{0}^{\infty} ds ~ \mu (s)~ g^*_{k,s}(z,\tau)  g_{k,s} (z',\tau') {1 \over  \mathsf{C}_{+-}}\;.\nonumber \\ \label{janusRL}
\end{align}
In the next two subsections we will evaluate this expression in two cases. First the undeformed solution with $c_J=0$ which corresponds to the $AdS_3$ vacuum and hence a trivial defect, and secondly for a nonzero Janus deformation parameter $c_J$ corresponding to a Janus interface.

\subsection{Case 1: $c_J = 0$}

When $c_J=0$, we have an exact  $AdS_3$  background and the connection coefficients become
\begin{equation}
 \mathsf{C}_{\epsilon\epsilon'}= {\Gamma(1-\epsilon(1-\Delta_-))\Gamma(\epsilon'(1-\Delta_-)) \over \Gamma((\epsilon'-\epsilon)){1-\Delta_- \over 2} +{1\over 2}\pm is)}\; ,
\end{equation}
which is a reflection of the fact that  Heun's equation simplifies to a hypergeometric equation for $c_J=0$.
The two-point functions are then given by
\begin{align}\label{cj0RR}
     \langle O(z,\tau) O(z', \tau')\rangle_{RR} =&\langle O(z,\tau) O(z', \tau')\rangle_{LL} \nonumber\\
     =&{(\Delta_+-\Delta_-) \Gamma(\Delta_--1) \sqrt{zz'} \over 2^{\Delta_+ +2} \pi^4 \Gamma(1-\Delta_-)} \int_{-\infty}^{\infty} dk \int_0^{\infty} ds ~s ~ \sinh(2\pi s) \nonumber\\
     &\Gamma({3\over 2}-\Delta_-\pm is) K_{is}(|k|z) K_{-is}(|k|z') e^{ik(\tau-\tau')}\;,
\end{align}
and
\begin{align}\label{cj0LR}
\langle O(z,\tau) O(z', \tau')\rangle_{RL/LR}=&{(\Delta_+-\Delta_-) \Gamma(\Delta_--1) \sin(\pi(\Delta_--1))\sqrt{zz'} \over 2^{\Delta_+ +1} \pi^4 \Gamma(1-\Delta_-)} \int_{-\infty}^{\infty} dk \int_0^{\infty} ds \; s   \nonumber\\
&\sinh(\pi s)\; \Gamma({3\over 2}-\Delta_-\pm is)K_{is}(|k|z) K_{-is}(|k|z') e^{ik(\tau-\tau')}\;.
\end{align}
The integral over the spectral parameter $s$ can be performed using the appropriate Kontorovich–Lebedev transforms
\begin{align}
    &\int_0^{\infty}x\sinh(\pi x)\Gamma(\lambda+ix)\Gamma(\lambda-ix)K_{ix}(\alpha)K_{ix}(\beta)dx & \nonumber \\
    &= 2^{\lambda-1}\pi^{3/2}\left({\alpha\beta\over \beta+\alpha}\right)^{\lambda}\Gamma({1\over 2}+\lambda)K_{\lambda}(\beta+\alpha), \quad \quad \quad  |\arg \alpha|<\pi, 0<\mathrm{Re} \lambda<{1\over 2}, \beta>0\;,
\end{align}
and
\begin{align}
    &\int_0^{\infty}x\sinh(2\pi x)\Gamma(\lambda+ix)\Gamma(\lambda-ix)K_{ix}(\alpha)K_{ix}(\beta)dx & \nonumber \\
    &= {2^{\lambda} \pi^{5/2}\over \Gamma({1\over 2}-\lambda)}\left({\alpha\beta\over |\beta-\alpha|}\right)^{\lambda}K_{\lambda}(|\beta-\alpha|)  \quad \quad \quad \alpha>0, 0<\mathrm{Re} \lambda<{1\over 2}, \beta>0,
\end{align}
which are equations (12) and (13) in Chapter 12 of \cite{ErdelyiV2}. Then (\ref{cj0RR}) and (\ref{cj0LR}) become
\begin{align}
    \langle O(z,\tau) O(z', \tau')\rangle_{RR/LL} =& {1-\Delta_- \over 2^{5/2}\pi^{3/2}} (\Delta_+ -\Delta_-)(zz')^{\Delta_+} \nonumber\\
     &\int_{-\infty}^{\infty} dk {1\over \Gamma(2-\Delta_-)} \left({|k| \over |z-z'|}\right)^{3/2-\Delta_-} K_{3/2-\Delta_-}(|k||z-z'|)e^{ik(\tau-\tau')}\;,\nonumber \\
     \langle O(z,\tau) O(z', \tau')\rangle_{RL/LR} =& {1-\Delta_- \over 2^{5/2}\pi^{3/2}} (\Delta_+ -\Delta_-)(zz')^{\Delta_+} \nonumber\\
     &\int_{-\infty}^{\infty} dk {1\over \Gamma(2-\Delta_-)} \left({|k| \over |z+z'|}\right)^{3/2-\Delta_-} K_{3/2-\Delta_-}(|k||z+z'|)e^{ik(\tau-\tau')}\;.\label{twopointcjzero}
\end{align}
The remaining integral over $k$ can by computed using the following Fourier transform  formula
\begin{align}
    {1\over \sqrt{2\pi} } \int dt e^{i \omega t} {1\over (a^2+t^2)^h} = {2^{1-h} \over \Gamma(h)} K_{h-{1\over 2}}(|a \omega|) \left( {|\omega|\over |a|}\right)^{h-{1\over 2}}.
\end{align}
Using this the two-point function (\ref{twopointcjzero}) becomes
\begin{align}
 \langle O(z,\tau) O(z', \tau')\rangle_{RR/LL} &={(\Delta_+-\Delta_-) (1-\Delta_-) (zz')^{\Delta_+} \over 2^{1+\Delta_-}\pi ((z-z')^2+(\tau-\tau')^2)^{\Delta_+}}\;,\\
 \langle O(z,\tau) O(z', \tau')\rangle_{RL/LR}&={(\Delta_+-\Delta_-) (1-\Delta_-) (zz')^{\Delta_+} \over 2^{1+\Delta_-}\pi ((z+z')^2+(\tau-\tau')^2)^{\Delta_+}}\;.
\end{align}
The two correlators are identical upon replacing $z'$ by $-z'$ in the $RL/LR$ correlators, and the $RR/LL$ correlators are those of a conformal primary of weight $\Delta_+$. This confirms that the $c_J=0$ case reproduces the two-point function of pure $AdS$ using the slightly unconventional $H_2$ slicing of $AdS_3$.
The $LR,RL$ correlator with insertions on either side of the interface is finite when the sources are brought together, unlike the $LL,RR$ case where both sources are on the same half space.

\subsection{Case 2: $c_J \neq 0$}
Unlike the  $AdS_3$ case, the integral over the spectral parameter $s$ is not readily computable when $c_J \neq 0$. We instead start by integrating the expressions (\ref{janusRL}) over $k$. Integrating the $H_2$ Laplacian eigenfunctions (see Appendix \ref{appendix:a}) over the momentum $k$ amounts to a cosine fourier transform of modified Bessel functions \cite{ErdelyiV1}
\begin{align}\int _0 ^{\infty} K_\nu (\alpha x) K_\nu (\beta x) \cos (xy) dx = {\pi^2 \over 4 \sqrt{\alpha \beta}} \sec(\pi \nu) P_{\nu-1/2} ({y^2+\alpha^2+\beta^2 \over 2\alpha \beta})\;.
\end{align}
This turns the two-points functions into the following expressions  
\begin{align}
    \langle O(z,\tau) O(z', \tau')\rangle = {(\Delta_+-\Delta_-) \over 4\pi} \int_{0}^{\infty} & ds ~s \tanh(\pi s) P_{\nu-1/2} ({(\tau-\tau')^2+z^2+z'^2 \over 2zz'}) \nonumber\\
    &\times [\text{suitable ratio of } \mathsf{C}]\;.
\end{align}
Using the hypergeometric representation of Legendre functions, we get an expression in terms of the cross-ratio $\xi = {(\tau-\tau')^2 + (z-z')^2 \over 4 zz'}$ which is denoted as $\xi_{LL,RR}$ in  (\ref{allxi})
\begin{equation}
\label{PF21}
    P_{is-1/2} ({(\tau-\tau')^2 + z^2 + z'^2 \over 2zz'}) = \;_{2}F_1 ({1 \over 2}+is,{1 \over 2}-is ; 1 ; -\xi)\;.
\end{equation}
The limit we are interested in is the case where the operator insertions are close to the defect but remain apart, meaning $\xi \rightarrow \infty$. Such an insertion admits a BOPE, whose relevant details were reviewed in Section \ref{sec:3.3}. To take this limit we use the appropriate transformation formula for the hypergeometric function, which is given by
\begin{align}
    \;_{2}F_1 ({1 \over 2}+is,{1 \over 2}-is ; 1 ; -\xi)&= {\Gamma(-2 is ) \over \Gamma({1\over 2}- i s)^2} (-\xi)^{-{1\over 2} - i s} \; _{2}F_1 ({1 \over 2}+is,{1 \over 2}+is ; 1+2is ; -{1\over \xi})  \nonumber\\
    &+ {\Gamma(2 is ) \over \Gamma({1\over 2}+ i s)^2} (-\xi)^{-{1\over 2} + i s} \; _{2}F_1 ({1 \over 2}-is,{1 \over 2}-is ; 1-2is ; -{1\over \xi}) \;.\label{hypgeotransf}
\end{align}
The two terms only differ by $s \rightarrow -s$, so we can extend the integral to [$-\infty,\infty$] and just keep one term. Plugging in the connection coefficients, we obtain the following expressions for the LL and RR two-point function
\begin{align}
     \langle O(z,\tau) O(z', \tau')\rangle_{RR} = {(\Delta_+-\Delta_-) \over 4\pi} \int_{-\infty}^{\infty}  &ds ~s \tanh(\pi s) {\Gamma(-2 is ) \over \Gamma({1\over 2}- i s)^2} (-\xi)^{-{1\over 2} - i s} \nonumber\\
     &\: _{2}F_1 ({1 \over 2}+is,{1 \over 2}+is ; 1+2is ; -{1\over \xi}) \nonumber\\
     &{\Gamma(\Delta_- -1) \Gamma({3\over 2}-\Delta_- \pm \sigma) \over \Gamma(1-\Delta_-)} {\cos(\pi \sigma) \over \pi} e^{-\partial_{\theta_1}\mathcal{W}(t)}\;, \label{wcoeffRR}\\
     \langle O(z,\tau) O(z', \tau')\rangle_{LL} = {(\Delta_+-Delta_-) \over 4\pi} \int_{-\infty}^{\infty}  &ds ~s \tanh(\pi s) {\Gamma(-2 is ) \over \Gamma({1\over 2}- i s)^2} (-\xi)^{-{1\over 2} - i s} \nonumber\\
     &\: _{2}F_1 ({1 \over 2}+is,{1 \over 2}+is ; 1+2is ; -{1\over \xi}) \nonumber\\
     &{\Gamma(\Delta_- -1) \Gamma({3\over 2}-\Delta_- \pm \sigma) \over \Gamma(1-\Delta_-)} {\cos(\pi \sigma) \over \pi} e^{-\partial_{\theta_0} \mathcal{W}(t)}\;. \label{wcoeffLL}
     \end{align}
For the LR/RL two-point function we utilize the following transformation formula
\begin{align}
    \;_{2}F_1 ({1 \over 2}+is,{1 \over 2}-is ; 1 ; -\xi)&= {\Gamma(-2 is ) \over \Gamma({1\over 2}- i s)^2} (1+\xi)^{-{1\over 2} - i s} \; _{2}F_1 ({1 \over 2}+is,{1 \over 2}+is ; 1+2is ; {1\over 1+\xi})  \nonumber\\
    &+ {\Gamma(2 is ) \over \Gamma({1\over 2}+ i s)^2} (1+\xi)^{-{1\over 2} + i s} \; _{2}F_1 ({1 \over 2}-is,{1 \over 2}-is ; 1-2is ; {1\over 1+\xi}) \;.\label{hypgeotransf2}
\end{align}
Since $\xi =\xi_{LL,RR}$ it follows from   (\ref{allxi}) that the combination  $1+\xi =\xi_{LR,RL}$  is the cross ratio for correlators across the interface. In the following we will denote $\xi_{LR,RL}=\tilde \xi$ for brevity.
\begin{align}
     \langle O(z,\tau) O(z', \tau')\rangle_{RL/LR} = {(\Delta_+-\Delta_-) \over 4\pi} \int_{-\infty}^{\infty}  &ds ~s \tanh(\pi s) {\Gamma(-2 is ) \over \Gamma({1\over 2}- i s)^2} (\tilde \xi)^{-{1\over 2} - i s}  \nonumber\\
     &\: _{2}F_1 ({1 \over 2}+is,{1 \over 2}+is ; 1+2is ; {1\over \tilde \xi}) \nonumber\\
     &{\Gamma({3\over 2}-\Delta_- \pm \sigma) \over \Gamma(2-\Delta_-) \Gamma(1-\Delta_-)} e^{-{1\over 2}(\partial_{\theta_1} + \partial_{\theta_0})\mathcal{W}(t)} \;.\label{wcoeffRL}
\end{align}
The integrands for the RR and LL correlators, which correspond to the insertion of both operators on the left and right, respectively, only differ by the exponent coming from the semi-classical block ${\cal W}$. The LR/RL correlators, which correspond to the insertion of the operators on two opposite sides have a different integrand compared to LL/RR. However, it is straightforward to see that the locations and orders of the poles are precisely the same for all three cases: gamma function poles from $\Gamma({3\over 2}-\Delta_- + \sigma)$ and $\Gamma({3\over 2}-\Delta_- - \sigma)$, both of order 1. 

We will evaluate the integral over the spectral parameter $s$ using the residue theorem, and the locations of the poles will be related to the conformal dimension of the boundary operators appearing in the BOPE due to the factors of $(-\xi)^{-{1\over 2} - i s}$ in (\ref{wcoeffRR}), (\ref{wcoeffLL}), and (\ref{wcoeffRL}),
see the discussion in Section \ref{sec:3.3}.
In selecting the first term in (\ref{hypgeotransf}), which has the form $\: _{2}F_1(\hat{\Delta},\hat{\Delta},2\hat{\Delta},-{1\over \xi})$, we've implicitly made a contour choice. The scaling dimensions of the boundary operators must obey $\hat{\Delta} \geq 0$ for a unitary CFT, so we will close the contour on the lower half plane, picking up contributions from the poles of $\Gamma({3\over 2}-\Delta_- - \sigma)$  
\begin{align}
\sigma_n= {3\over 2}-\Delta_- +n ,\quad\quad n=0,1,2,\cdots\;.
\end{align}
We can invert the relation (\ref{sigmas}) relating $\sigma$ and $s$ for  shifted pole locations perturbatively up to order $c_J^2$ and obtain
\begin{align}\label{snlocation}
    s_n = &-i({3\over 2}+n-\Delta_-) \nonumber \\
    &+{i\over 128}\big(32(n+1)+{n(n-1)\over 1/2+n-\Delta_-}+{2-10n(n+1)\over 3/2+n-\Delta_-}+{(n+1)(n+2)\over 5/2+n-\Delta_-} -16\Delta_- \big) c_J^2 \nonumber \\
    &+\mathcal{O}(c_J^3)\;.
\end{align}
As discussed above, comparing (\ref{wcoeffRR}), (\ref{wcoeffLL}), and (\ref{wcoeffRL}) to (\ref{bope}), the  poles give us the the boundary operator scaling dimension
\begin{equation}
    \hat{\Delta}_n={1\over2}+is_n\;.
\end{equation}
This is one of our main results.
The Janus deformation hence induces a shift in the boundary operator dimension of second order in  $c_J$.
The factor of $|zz'|^{-\Delta_+}$ in (\ref{2ptitocr}) can be restored by going to a flat conformal frame. The BOPE coefficients $b_{\mathcal{O}\hat{\mathcal{O}}_n}$ are then the residues evaluated at the shifted poles.

\begin{align}
      \langle O(z,\tau) O(z', \tau')\rangle_{\small L/R,L/R} &= \sum_{\hat \Delta_n} b_{\mathcal{O}_{L/R}\hat{\mathcal{O}}_n}b_{\mathcal{O}_{L/R}\hat{\mathcal{O}}_n} (-\xi)^{-\hat \Delta_n}\: _{2}F_1 (\hat \Delta_n,\hat \Delta_n; 2\hat \Delta_n; -{1\over \xi}) \;,\nonumber\\
       \langle O(z,\tau) O(z', \tau')\rangle_{\small L/R,R/L} &= \sum_{\hat \Delta_n} b_{\mathcal{O}_{L/R}\hat{\mathcal{O}}_n}b_{\mathcal{O}_{R/L}\hat{\mathcal{O}}_n} (\tilde\xi)^{-\hat \Delta_n}\: _{2}F_1 (\hat \Delta_n,\hat \Delta_n; 2\hat \Delta_n; {1\over \tilde\xi}) \;.
\end{align}
Here the sum is over the boundary operators $\hat{\mathcal{O}}_n$  which are labeled by positive integers $n$.
Evaluating the residues of the integrals (\ref{wcoeffRR}), (\ref{wcoeffLL}), and (\ref{wcoeffRL})  allows us to read off the BOPE coefficients for the $LL,RR,LR$, and $RL$  correlators
\begin{align}\label{bopea}
b_{\mathcal{O}_L\hat{\mathcal{O}}_n}b_{\mathcal{O}_L\hat{\mathcal{O}}_n} &=   {(\Delta_+-\Delta_-) \over 2} s_n \tanh(\pi s_n) {\Gamma(-2 is_n ) \over \Gamma({1\over 2}- i s_n)^2} {\Gamma(n+3-2\Delta_-) \over  \Gamma(1-\Delta_-)\Gamma(2-\Delta_-) n!}e^{-\partial_{\theta_1}\mathcal{W}\mid_{s=s_n}}\;,\\
\label{bopeb}
b_{\mathcal{O}_R\hat{\mathcal{O}}_n}b_{\mathcal{O}_R\hat{\mathcal{O}}_n} &=   {(\Delta_+-\Delta_-) \over 2} s_n \tanh(\pi s_n) {\Gamma(-2 is_n ) \over \Gamma({1\over 2}- i s_n)^2} {\Gamma(n+3-2\Delta_-) \over  \Gamma(1-\Delta_-)\Gamma(2-\Delta_-) n!}e^{-\partial_{\theta_0}\mathcal{W}\mid_{s=s_n}}\;,\\
\label{bopec}
b_{\mathcal{O}_L\hat{\mathcal{O}}_n}b_{\mathcal{O}_R\hat{\mathcal{O}}_n} &=   {(\Delta_+-\Delta_-) \over 2} s_n \tanh(\pi s_n) {\Gamma(-2 is_n ) \over \Gamma({1\over 2}- i s_n)^2} {(-1)^n\Gamma(n+3-2\Delta_-) \over  \Gamma(1-\Delta_-)\Gamma(2-\Delta_-) n!}\nonumber\\
&\;\;\;\;\; \times e^{-{1\over 2} \partial_{\theta_0}\mathcal{W}\mid_{s=s_n}-{1\over 2}\partial_{\theta_1}\mathcal{W}\mid_{s=s_n} }\;.
\end{align}
Where the residues are evaluated at the shifted locations (\ref{snlocation}) using the formula 
\begin{align}
    \Gamma(1-z)\Gamma(z)= {\pi \over \sin \pi z}\;.
\end{align}
Note that the BOPE coefficients satisfy a consistency relation where the product of (\ref{bopea}) and (\ref{bopeb}) is equal to the square of (\ref{bopec}).

\section{Discussion} \label{sec:6}
In this paper, we studied a new occurrence of  Heun's equation in theoretical physics, namely  the equation of motion of a massive probe scalar in an $AdS_3$ Janus solution. The two-point functions were calculated using the connection coefficients of  Heun's equation, obtained by taking advantage of degenerate Virasoro blocks in a Liouville CFT. We identified the shift in conformal dimension of the boundary CFT operators dual to the probe scalar field, and also found the BOPE coefficients with different operator insertions with respect to the Janus defect. These results are given as a expansion for small Janus deformation parameter $c_J$. We presented the first nontrivial order which is quadratic in $c_J$. It is in principle possible to obtain higher orders using the expansion of the semi-classical conformal blocks, but we did not present these as the expressions become very unwieldy.  

For generic mass of the probe scalar the two Frobenius solutions are linearly independent. For special mass values where $\Delta_+ -\Delta_-$ is an integer, the solutions to the equation of motion become degenerate, i.e. the index equation gives two identical roots. We did not discuss this case in the present paper but the resulting  logarithmic correction can be handled using methods discussed recently in \cite{jia2024}.

The three-dimensional Euclidean Janus solution can be  used to construct wormhole solutions by taking a quotient of the Euclidean $H_2$ slice by a discrete Fuchsian group $\Gamma$, yielding a compact Riemann surface $\Sigma_2= H_2/\Gamma$. The resulting Riemann surface has genus $g\geq2$ and the wormhole would relate two CFTs defined on $\Sigma$ with different marginal deformations. Note that the  boundary of the Euclidean AdS space $H_2$ is removed by the quotient and hence operators localized on opposite ends of the wormhole cannot come close to each other, while operators on the same side can still have a bulk OPE.
For compact $H_2/\Gamma$ the spectrum of the two-dimensional Laplacian will be discrete. This means that in the expressions for the correlators in Section \ref{sec:5}, the integrals over $s$ will be replaced by discrete sums.
While obtaining the spectrum is a hard mathematical problem, it may still be possible to consider the bulk OPE channel which is dominated by small distances, hence the sum over $s$ may be approximated by an integral.

In this paper we considered a massive scalar in a simple Janus background in the probe approximation, i.e. we neglected any backreaction of the probe scalar on the Janus solution as well as any other couplings to other fields which may be present.  It would be interesting to apply the methods presented in this paper to supergravity theories which are consistent truncations of ten-dimensional solutions, such as gauged $d=3, N=8$ gauged supergravity \cite{Nicolai:2001ac}. Supersymmetric Janus and RG-flow interface solutions have been constructed for such theories, see e.g. \cite{Chen:2020efh,Chen:2021mtn,Gutperle:2022fma}. This may be interesting for two reasons: First, even in a probe approximation, couplings to other background fields may modify the linearized equation for the massive scalar field which may not be of Heun form at all (or correspond to confluent limits). Second, there have been several interesting relations and inequalities of holographic observables in holographic interface theories involving various forms of entanglement entropy  and reflection and transmission coefficients \cite{Karch:2023evr,Karch:2024udk,Baig:2024hfc}. It may be possible that the two-point functions which have been calculated here could also obey such relations and provide useful data in describing holographic interfaces. 
We leave these interesting questions for future research.

\acknowledgments
The work of M.G. was supported, in part, by the National Science Foundation under grant PHY-2209700. 
Both authors  are grateful to the Bhaumik Institute for support.

\newpage

\appendix
\section{Eigenfunctions of the Laplacian on $H_2$}\label{appendix:a}

We are looking for an orthogonal basis of eigenfunctions of the Laplacian $\nabla^2_{H_2}$ on the upper half plane ($\zeta = \tau + iz, ~z>0$). 
The eigenfunctions of the two-dimensional Laplacian on $H_2$ satisfy
\begin{align}
 \nabla^2_{H_2} g_s  = z^2 (\partial_z^2+\partial_\tau^2) g_s = -(s^2+{1\over 4} )g_s\;,
\end{align}
and we have
\begin{align} 
g_{k,s} = e^{i k \tau } z^{1\over 2} K_{is}(|k| z)\;,
\end{align}
where $K_\nu$ is the modified Bessel (or MacDonald) function.
One completeness/orthogonality relation is given by
\begin{align}\label{ortho1}
     \int {dz d\tau\over z^2} \mu(s) g_{k,s}^*(z, \tau)  g_{k',s'}(z,\tau)  &= \int {dz d\tau\over z^2} {s \sinh(\pi s) \over \pi^3} g_{k,s}^*(z, \tau)  g_{k',s'}(z,\tau) \nonumber \\
     &= \delta(k-k')\delta(s-s')\;.
\end{align}
The second completeness/orthogonality relation is obtained by integrating
\begin{align}
    {2\over \pi^2} \int_0^\infty ds\; s \sinh(\pi s) K_{is}(x) K_{is}(x') = x\delta(x-x')\;,
\end{align}
using the Kontorovich-Lebedev transform pair \cite{Stone}
\begin{align}
\tilde f(s) &= \int_0^\infty dx K_{is}(x) f(x) \nonumber \\
f(x) &= {1\over \pi^2 x } \int_0^ \infty  ds\; 2 s \sinh (\pi s) K_{i s} (x)\tilde f(s)\;,
\end{align}
which, together with the integral over k, leaves us with
\begin{align}\label{measure}
  {1\over \pi^3} \int_{-\infty}^\infty dk   \int_0^\infty ds   s \sinh (\pi s) g^*_{k,s}(z,\tau)  g_{k,s} (z',\tau') &= z^2 \delta(z-z')\delta(\tau-\tau')\;.
\end{align}

\subsection{Boundedness of spectrum of $\nabla^2_{H_2}$ }

There is an argument using the Cauchy-Schwarz inequality to show that the eigenvalues of the Laplacian on $H_2$ lie in $[-\infty, -{1\over 4}]$.
We'll start with the integral
\begin{align}
    I= {1\over 4} \Big( \int_0^\infty f^2 {1\over z^2}dz\Big)^2\;.
\end{align}
We can integrate the identity 
\begin{align}
    \partial_z\big(f^2 {1\over z}\big) = 2 \partial_z f f{1\over z} -f^2{1\over z^2}\;,
\end{align}
where the integral on the left hand side will vanish, following from square integrability. Then we have
\begin{align}
    I = \Big( \int_0^\infty \partial_z f f {1\over z} dz\Big)^2 \;.
\end{align}
We can now use the Cauchy-Schwarz inequality $|(u,v)|^2 \leq (u,u) (v,v)$ where the inner product is defined as the integral over $z$. This gives the inequality
\begin{align}
    I\leq  \int_0^\infty (\partial_zf)^2 dz \int_0^\infty f^2 {1\over z^2}
dz\;.
\end{align}
The second factor on the right hand side is one factor in the square on the left hand side, so dividing by it gives
\begin{align}
    {1\over 4}\int_0^\infty f^2 {1\over z^2} dz\leq \int_0^\infty (\partial_zf)^2 dz\;.
\end{align}
Since this inequality is true point-wise, we can make everything dependent on $\tau$ and integrate over $\tau\in [-\infty,+\infty]$.
We can also add the positive definite term $\int (\partial_\tau f)^2$ to the right hand side without violating the inequality. Finally we multiply by $1={z^2 \over z^2}$ inside the integral. Putting this all together we get
\begin{align}
    {1\over 4}\int_{H_2} f^2 {dz d\tau \over z^2}    \leq  \int_{H_2} z^2 \Big( (\partial_z f)^2 +(\partial_\tau f)^2\Big) {dz d\tau \over z^2}\;.
\end{align}
Integrating the right hand side by parts for both $z,\tau$ gives
\begin{align}
    {1\over 4}\int_{H_2} f^2 {dz d\tau \over z^2}    &\leq - \int_{H_2} f  z^2 \Big( \partial_z^2 +\partial_\tau^2 \big) f {dz d\tau \over z^2} \nonumber\\
    &\leq - \int_{H_2} f  \nabla^2 f  {dz d\tau \over z^2}\;.
\end{align}
If $f$ is an eigenfunction of the Laplacian, $ \nabla^2 f=\lambda f$ and we get 
\begin{align}
    \lambda \leq -{1\over 4}\;.
\end{align}
Writing $\lambda = \nu(\nu+1) $ 
with $\nu= i s-1/2 $ we find that $\lambda = -s^2-{1\over 4} \leq -{1\over 4}$, meaning $s$ must be real-valued.

\section{Bulk and bulk-boundary Green's functions}\label{appendix:b}

Green's functions for linear second order ODEs can be constructed using the  the Wronskian method\footnote{See \cite{estes,Melby-Thompson:2017aip} for related constructions of Green's functions in holography.}. The Wronskian for an ODE is defined as 
\begin{align}
W(\phi_1,\phi_2)(y)= \phi_1 (y) \partial_y \phi_2- \phi_2 (y) \partial_y \phi_1(y)\;,
\end{align}
where $\phi_{1,2}$ are two linearly independent solutions of the ODE. We work with the normal form of Heun's equation (\ref{probeHeun}) rather the one coming from the probe scalar (\ref{EOM_h}), since
\begin{align}\label{wronski}
    \partial_y W = \phi_1 (y) \partial^2_y \phi_2- \phi_2 (y) \partial^2_y \phi_1(y)
\end{align}
vanishes for the normal form meaning the corresponding Wronskian is constant.  
The bulk Green's function is constructed by choosing $\phi_{1,2}$ to be $N_s^{L/R}$, defined in (\ref{normsoln}):
\begin{align}
  G(y, z, \tau|y', z', \tau') =& {1\over \pi^3} \int_{-\infty}^\infty dk   \int_0^\infty ds ~ s ~ \sinh (\pi s)  g_{k,s}^*(z,\tau) g_{k,s} (z',\tau') \nonumber \\
  &\frac{\theta(y-y') N_s^L(y') N_s^R(y) + \theta(y'-y) N_s^L(y) N_s^R(y')}{W(N_s^L,N_s^R)(y)}\; .
\end{align}
From (\ref{normsoln}) and (\ref{nnormsoln}), we can confirm that the Wronksian is constant and given by
\begin{align}
    W(N_s^L,N_s^R)(y) = {\mathsf{C}_{+-} \over 2} (\Delta_+ - \Delta_-)\;.
\end{align}
The bulk-boundary propagators with sources on either boundary are then given by
\begin{align}\label{khat}
\hat K_R(y, z, \tau|z', \tau')&={\Delta_+ - \Delta_- \over 2} \lim_{y'\to 1} {1\over (1-y')^{\Delta_+ /2}} G(y, z, \tau|y', z', \tau')\;, \\
\hat K_L(y, z, \tau|z', \tau')&={\Delta_+ - \Delta_- \over 2} \lim_{y'\to 0} {1\over y'^{\Delta_+ /2}} G(y, z, \tau|y', z', \tau')\;,
\end{align}
with the normalizations fixed so that
\begin{align}
    \Big( \partial_y^2  + T(y)\Big)    \hat K_{L,R} (y, z, \tau| z',\tau')&=0\;,  \nonumber\\
    \lim_{y\to 1} \hat K_R (y, z, \tau| z',\tau') &= (1-y)^{\Delta_-/2} z^2 \delta(z-z')\delta(\tau-\tau')\;,\nonumber \\
    \lim_{y\to 0} \hat K_R (y, z, \tau| z',\tau') &= 0\;,\nonumber \\
      \lim_{y\to 0} \hat K_L (y, z, \tau| z',\tau') &= y^{\Delta_-/2} z^2 \delta(z-z')\delta(\tau-\tau')\;,\nonumber \\
    \lim_{y\to 1} \hat K_L (y, z, \tau| z',\tau') &= 0\;.
\end{align}
The bulk-boundary Green's function for the original probe scalar can be obtained by rescaling  $\hat K_{L/R}$ by  the factor (\ref{solnresc})
\begin{align} \label{bulkbndyprop}
    K_R(y, z, \tau| z',\tau')&= {1\over \pi^3 \sqrt{1-c_J+2 c_Jy }} \int_{-\infty}^\infty dk   \int_0^\infty ds ~ s ~ \sinh (\pi s)  g_{k,s}^*(z,\tau) g_{k,s} (z',\tau')  {N_s^L(y)\over \mathsf{C}_{+-}} \;,\nonumber\\
     K_L(y, z, \tau| z',\tau')&= {1\over \pi^3 \sqrt{1-c_J+2 c_Jy}} \int_{-\infty}^\infty dk   \int_0^\infty ds ~ s ~ \sinh (\pi s)  g_{k,s}^*(z,\tau) g_{k,s} (z',\tau')  {N_s^R(y)\over \mathsf{C}_{+-}}\;.
\end{align}
The bulk field for the probe scalar with sources on the left and the right boundary can then be written as
\begin{align} \label{phiwsources}
    \psi(y, z, \tau)&= \int {dz'd\tau'\over (z')^2} \Big(  K_R(y, z, \tau| z',\tau') J_R(z',\tau') +   K_L(y, z, \tau| z',\tau') J_L(z',\tau')\Big)\;.
\end{align}

\section{Heun functions}\label{appendix:c}
In this appendix we review some properties of Heun's differential equation and its local Frobenius solutions, following the conventions found in Section 31 of \cite{NIST:DLMF}.  
Heun's equation (\ref{probeHeun}) is a 2nd order Fuchsian differential equation with 4 regular singularities ($z=0,1,a,\infty$)

\begin{align} \label{heuneq}
H''+\Big( {\gamma\over z}+ {\delta \over z-1} + {1+\alpha+\beta-\gamma-\delta \over z-a}\Big) H'+{\alpha \beta z-q\over z(z-1)(z-a)}H=0\;.
\end{align}
Near the critical point  $z=0$ the Frobenius solution 
that evaluates to 1 at $z=0$  has the power series expansion
\begin{align}
 H \ell(a, q ; \alpha, \beta, \gamma, \delta ; z)=\sum_{j=0}^{\infty} A_j z^j   \;.
\end{align} The Frobenius coefficients $A_i$ satisfy the recursion relations
\begin{align}
&A_0=1\;,\nonumber\\
&A_1={q\over a \gamma}A_0\; \nonumber\\
&R_j A_{j+1}-\left(Q_j+q\right) A_j+P_j A_{j-1}=0\;,
\end{align}
with
\begin{align}
P_j & =(j-1+\alpha)(j-1+\beta)\;, \nonumber\\
Q_j & =j((j-1+\gamma)(1+a)+a\;,\nonumber \delta+1+\alpha+\beta-\gamma-\delta), \\
R_j & =a(j+1)(j+\gamma)\; .
\end{align}
The other solution at $z=0$ (for $\gamma \neq 1,2,3, \cdots$) is
\begin{equation}
    z^{1-\gamma}H\ell\begin{pmatrix}a,(a\delta+\varepsilon)(1-\gamma)+q;\alpha+1-\gamma,\beta+1-\gamma,2-\gamma,\delta;z\end{pmatrix}\;,
\end{equation}
and the solutions at $z=1$ (with similar restrictions on $\delta$) are
\begin{align}
    &H\ell\left(1-a,\alpha\beta-q;\alpha,\beta,\delta,\gamma;1-z\right),(1-z)^{1-\delta}\;,\nonumber\\
    &H\ell\left(1-a,((1-a)\gamma+\varepsilon)(1-\delta)+\alpha\beta-q;\alpha+1-\delta,\beta+1-\delta,2-\delta,\gamma;1-z\right)\;.
\end{align}
Subtleties arise when $\gamma, \delta=1,2,3,\cdots$; such logarithmic cases are also discussed in \cite{NIST:DLMF}. We omit that discussion here as this does not become an issue for the solutions we handle in this paper, as long as the dilaton mass is kept generic.

\medskip

\section{Virasoro conformal blocks} \label{appendix:d}
The connection coefficients that relate the probe scalar solutions near either AdS boundary come from fusion rules for degenerate Virasoro blocks in a Liouville CFT. This CFT is distinct from the $\text{CFT}_2$ that is dual to $\text{AdS}_3$. The s-channel connection formulae are considerably simpler than the traditional t-channel result. What follows is a summary of the findings in \cite{jia2024} and \cite{Lisovyy:2022flm}.

The central charge of a 2D Liouville CFT can be parametrized by the background charge $Q$
\begin{equation}
c=1+6Q^2, ~ Q=b+{1\over b}\;,
\end{equation}
where $b$ can be thought of as a coupling constant (which we will later take to 0). Primary operators in this CFT are labeled by their Liouville momenta $P$, and have conformal weight
\begin{equation}
    h={Q^2 \over 4}- P^2\;.
\end{equation}
Degenerate Verma modules occur for a discrete set of Liouville momenta, parametrized by two integers $r$ and $b$
\begin{equation}
    P_{<r,s>}={1\over 2} (rb+{s\over b})\;.
\end{equation}
The s-channel conformal blocks, which we get from inserting the OPE in a four-point function of primary operators, take on a series expansion in terms of the cross-ratio $t$, with coefficients fixed from instanton counting using the AGT relation \cite{Alday:2009aq}.

When we insert a degenerate primary into the four-point conformal blocks, the resulting five-point conformal blocks obey the BPZ equation with different insertions related by the fusion coefficients. The insertions we are interested in are
\begin{figure}[H] \label{fig:5pt}
\centering
\begin{subfigure}[b]{0.55\textwidth}
   \includegraphics[width=1\linewidth]{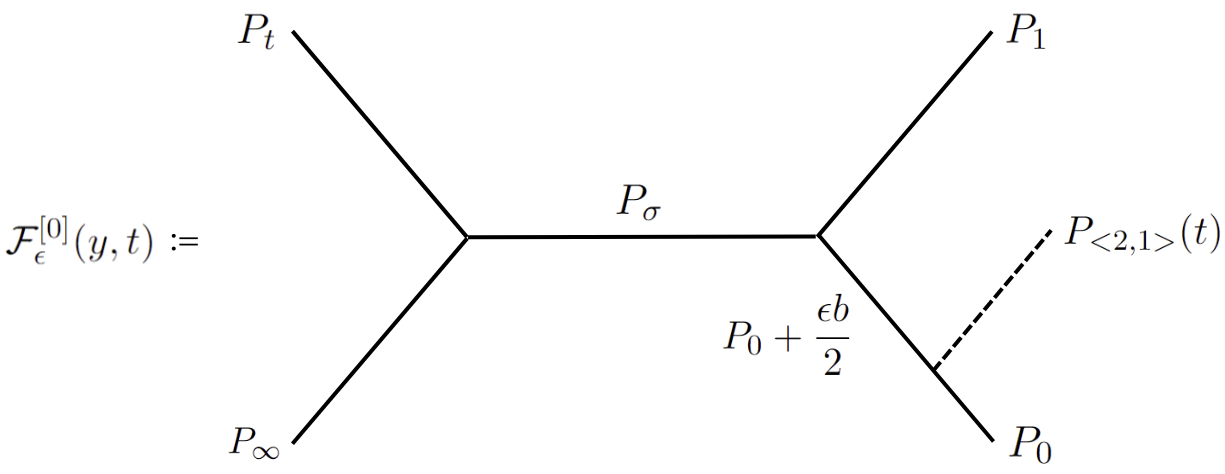}
\end{subfigure}

\vspace{5mm}

\begin{subfigure}[b]{0.55\textwidth}
   \includegraphics[width=1\linewidth]{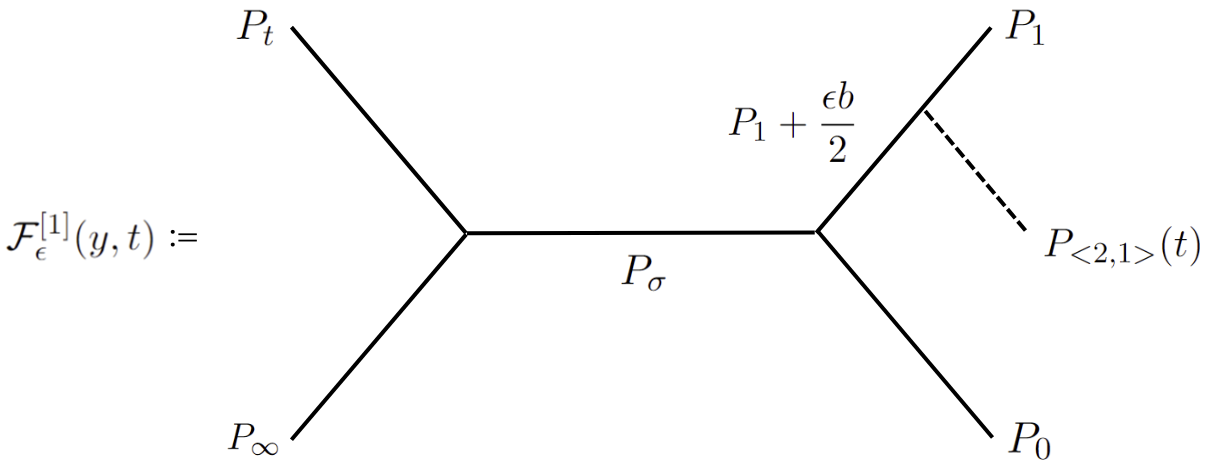}
\end{subfigure}
\end{figure}
\noindent with the intermediate momenta coming from degenerate-primary fusion rules.

Respectively taking the $y\rightarrow 0$ and $y\rightarrow 1$ OPE limits, the leading behavior becomes
\begin{figure}[H]
\centering
\includegraphics[height=0.45\textwidth]{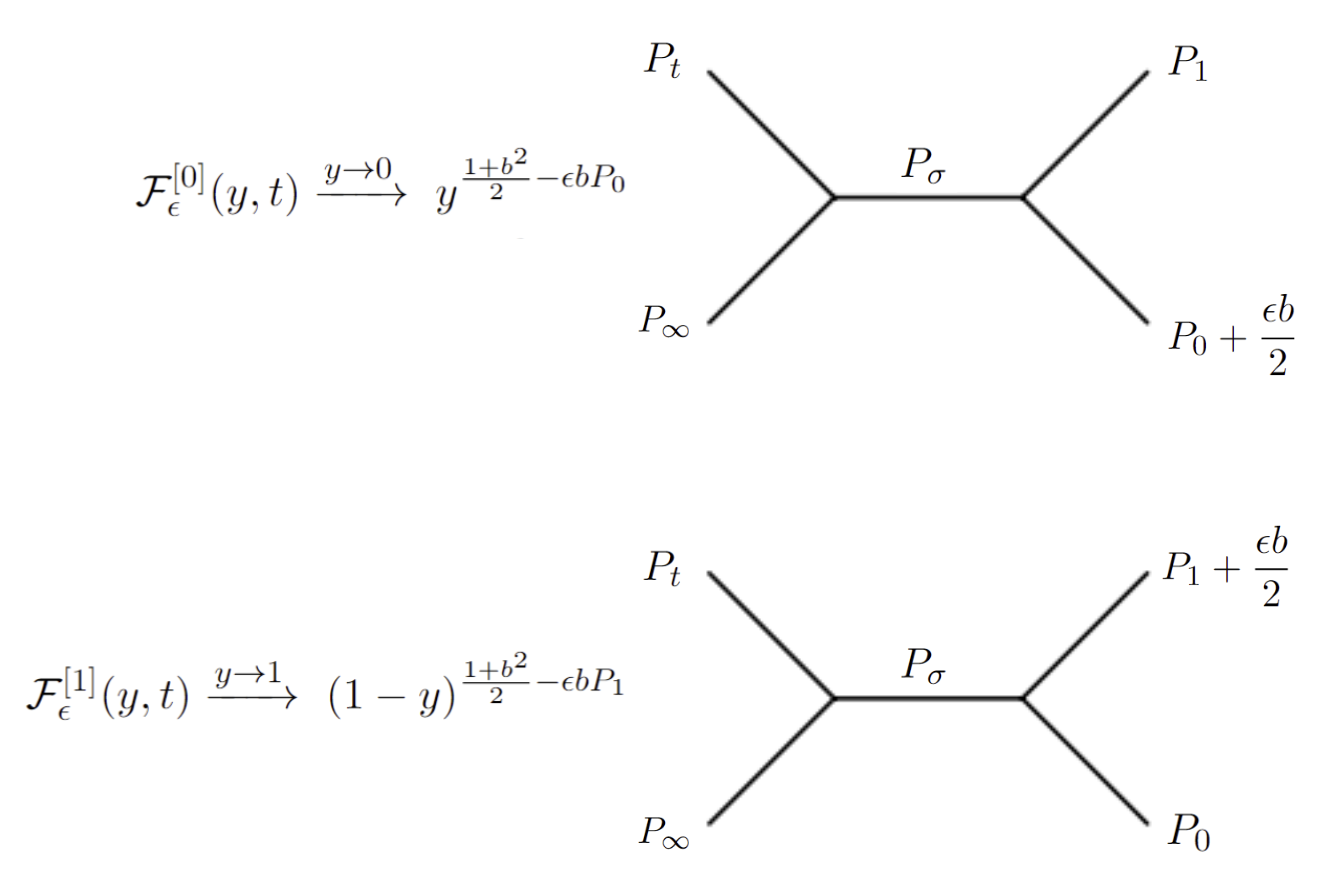}
\caption{Degenerate 4-point blocks with the degenerate operator fusing with the y=0 and y=1 punctures.}
\label{fig:4pt}
\end{figure}
\noindent and the leading exponents will later become the critical exponents for the relevant Frobenius solutions to Heun's equation.

The BPZ equation for the degenerate four-point blocks is
\begin{equation} \label{BPZ}
    ({1\over b^2}\partial_y^2+\mathcal{L}_{-2})\mathcal{F}(y,t)=0\;,
\end{equation}
with $\mathcal{L}_{-2}$ the differential operator
\begin{align}
    \mathcal{L}_{-2}=&\frac{h_{0}}{y^{2}}+\frac{h_{1}}{(y-1)^{2}}+\frac{h_{t}}{(y-t)^{2}}+\frac{h_{\infty}-h_{\langle2,1\rangle}-h_{0}-h_{1}-h_{t}}{y\left(y-1\right)} \nonumber\\
    &+\frac{t\left(t-1\right)}{y\left(y-1\right)\left(y-t\right)}\partial_{t}-\left(\frac{1}{y}+\frac{1}{y-1}\right)\partial_{y}\;.
\end{align}
The degenerate blocks in Figure \ref{fig:4pt} are related by a fusion transformation matrix
\begin{equation}
    \mathcal{F}^{[0]}_\epsilon (y,t)=\sum_{\epsilon'=\pm}\mathsf{F}_{\epsilon \epsilon'}(P_0,P_1,P_{\sigma})\mathcal {F}^{[1]}_\epsilon (y,t)\;,
\end{equation}
with
\begin{equation}
    \mathsf{F}_{\epsilon \epsilon'}(P_0,P_1,P_{\sigma})={\Gamma(1-2\epsilon b P_0) \Gamma(2\epsilon ' b P_1) \over \Gamma({1\over2}-\epsilon b P_0 + \epsilon ' b P_1 \pm b P_\sigma)}\;.
\end{equation}

When we take the semiclassical limit, i.e.
\begin{equation}
b\rightarrow 0, \quad P_i, P_\sigma \rightarrow \infty,\quad bP_i \rightarrow \theta_i, \quad bP_\sigma \rightarrow \sigma\;,
\end{equation}
we get several simplifications. The conformal weights of each operator factorize into the critical exponents
\begin{equation}
    h_i \rightarrow \delta_i = ({1\over 2}+\theta_i)({1\over 2}-\theta_i)\;,
\end{equation}
and more importantly the conformal blocks factorize into the heavy ($\Delta_i=\mathcal{O}(c^1)$, corresponding to the punctures) and light ($\Delta_i=\mathcal{O}(c^0)$, corresponding to the degenerate operator insertion) parts. The heavy block exponentiates, as proven in \cite{Be_ken_2020}, leaving
\begin{equation} \label{semiclass}
    \mathcal{F}_{\epsilon}^{[i]}(y,t) \xrightarrow{\text{semiclassical}} \mathcal{N}_{\epsilon}^{[i]}\rho_{\epsilon}^{[i]}(y)\exp\left[b^{-2}\mathcal{W}(t)\right],\quad i=0,1\;.
\end{equation}
$\mathcal{N}_{\epsilon}^{[i]}=\exp[{\epsilon\over 2}\partial_{\theta_i}\mathcal{W}(t)]$ is a normalization factor that accounts for the shift in conformal weights coming from fusion rules, and $\mathcal{W}(t)$ is the classical s-channel Virasoro block
\begin{figure}[H]
\centering
\includegraphics[width=0.4\textwidth]{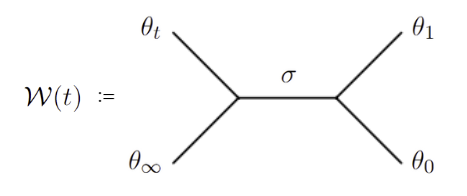}
\label{fig:classical}
\end{figure}
Plugging (\ref{semiclass}) back into the BPZ equation (\ref{BPZ}) and solving for $\rho_{\epsilon}^{[i]}(y)$, we finally get the connection coefficients (\ref{trieste}) for the normal form of  Heun's equation.

\section{Agreement between the Trieste and Sch\"afke-Schmidt Formulae} \label{appendix:e}

The Sch\"afke-Schmidt formula for the normal form of  Heun's equation \cite{Lisovyy:2022flm} is given by
\begin{equation}
\mathsf{C}_{\epsilon\epsilon'}^{SS}=\frac{\Gamma\left(1-2\epsilon \theta_0\right)\Gamma\left(2\epsilon' \theta_1\right)\left(1-{1\over t} \right)^{-\frac{1}{2}-\theta_t}}{\Gamma\left(\Theta_{10} \pm \omega\right)}\exp\sum_{k=1}^{\infty}\ln\left(1-{1\over t}\alpha_{k-1}-\frac{{1\over t}\beta_{k}}{1-{1\over t}\alpha_{k}-\frac{{1\over t}\beta_{k+1}}{1-...}}\right)\;,
\end{equation}
With $\Theta_{10}={1\over 2} +\epsilon' \theta_1-\epsilon \theta_0$ defined for brevity and
\begin{align}
 & \alpha_{k}=-\frac{\left(k+\frac{1}{2}-\epsilon\theta_{0}-\theta_{t}\right)^{2}-\theta_{0}^{2}-\theta_{\infty}^{2}+\omega^{2}}{\left(k+\frac{1}{2}-\epsilon\theta_{0}+\epsilon'\theta_{1}\right)^{2}-\omega^{2}}\;, \nonumber\\
 & \beta_{k}=\frac{k\left(k-2\epsilon\theta_0\right)\left(\left(k-\epsilon\theta_0+\epsilon'\theta_1-\theta_t\right)^2-\theta_\infty^2\right)}{\left(\left(k+\frac{1}{2}-\epsilon\theta_0+\epsilon'\theta_1\right)^2-\omega^2\right)\left(\left(k-\frac{1}{2}-\epsilon\theta_0+\epsilon'\theta_1\right)^2-\omega^2\right)}\;.
\end{align}
Truncating the infinite fraction to order ${1\over t}$ then expanding the logs for small ${1\over t}$, we have
\begin{equation}
    \mathsf{C}_{\epsilon\epsilon'}^{SS} \approx \frac{\Gamma\left(1-2\epsilon \theta_0\right)\Gamma\left(2\epsilon' \theta_1\right)}{\Gamma\left(\Theta_{10} \pm \omega\right)}\exp [({1\over 2} + \theta_t-\sum_{k=1}^{\infty}(\alpha_{k-1}+\beta_k)){1\over t}]\;.
\end{equation}
Resumming the exponent in terms of polygamma functions, we are left with 
\begin{align}
   \mathsf{C}_{\epsilon\epsilon'}^{SS}\approx &\frac{\Gamma\left(1-2\epsilon \theta_0\right)\Gamma\left(2\epsilon' \theta_1\right)}{\Gamma\left(\Theta_{10} \pm \omega\right)} \nonumber\\
   &\exp[-\big(\frac{(\epsilon\theta_0+\epsilon'\theta_1)\left(\frac{1}{4}-\omega^2+\theta_\infty^2-\theta_t^2\right)}{2\left(\frac{1}{4}-\omega^2\right)} + (\psi(\Theta_{10}+\omega)-\psi(\Theta_{10}-\omega))\sigma_1 \big) {1\over t}] \nonumber\\
   \approx &\frac{\Gamma\left(1-2\epsilon \theta_0\right)\Gamma\left(2\epsilon' \theta_1\right)}{\Gamma\left(\Theta_{10} \pm \omega\right)}\exp[-\frac{(\epsilon\theta_0+\epsilon'\theta_1)\left(\frac{1}{4}-\omega^2+\theta_\infty^2-\theta_t^2\right)}{2\left(\frac{1}{4}-\omega^2\right)}{1\over t}] \nonumber\\
   &[1-\big(\psi\left(\Theta_{10}+\omega\right)-\psi\left(\Theta_{10}-\omega\right)\big){\sigma_1 \over t}]\;.
\end{align}
Expanding the Trieste formula in ${1\over t}$, we have
\begin{align}
    \mathsf{C}_{\epsilon\epsilon'}^{T} \approx &\frac{\Gamma\left(1-2\epsilon \theta_0\right)\Gamma\left(2\epsilon' \theta_1\right)}{\Gamma\left(\Theta_{10} \pm (\omega+{\sigma_1\over t})\right)}\exp[-\frac{(\epsilon\theta_0+\epsilon'\theta_1)\left(\frac{1}{4}-\omega^2+\theta_\infty^2-\theta_t^2\right)}{2\left(\frac{1}{4}-\omega^2\right)} {1\over t}] \nonumber\\
    \approx & \Gamma\left(1-2\epsilon \theta_0\right)\Gamma\left(2\epsilon' \theta_1\right)\exp[-\frac{(\epsilon\theta_0+\epsilon'\theta_1)\left(\frac{1}{4}-\omega^2+\theta_\infty^2-\theta_t^2\right)}{2\left(\frac{1}{4}-\omega^2\right)} {1\over t}] \nonumber\\
    & [{1\over \Gamma\left(\Theta_{10} \pm \omega\right)}-\big( {\Gamma'(\Theta_{10}+\omega) \over \Gamma^2(\Theta_{10}+\omega)}{1 \over \Gamma(\Theta_{10}-\omega)}-{1 \over \Gamma(\Theta_{10}+\omega)}{\Gamma'(\Theta_{10}-\omega) \over \Gamma^2(\Theta_{10}-\omega)}\big){\sigma_1 \over t}] \nonumber\\
    = & \frac{\Gamma\left(1-2\epsilon \theta_0\right)\Gamma\left(2\epsilon' \theta_1\right)}{\Gamma\left(\Theta_{10} \pm \omega\right)}\exp[-\frac{(\epsilon\theta_0+\epsilon'\theta_1)\left(\frac{1}{4}-\omega^2+\theta_\infty^2-\theta_t^2\right)}{2\left(\frac{1}{4}-\omega^2\right)} {1\over t}] \nonumber\\
    & [1-\big(\psi\left(\Theta_{10}+\omega\right)-\psi\left(\Theta_{10}-\omega\right)\big){\sigma_1 \over t}]\;.
\end{align}
We've only shown explicit agreement to order ${1\over t}$, but the two formulae exactly agree to all orders; see Table 1 in \cite{Ren:2024ifh}.

\newpage

\providecommand{\href}[2]{#2}\begingroup\raggedright\endgroup

\end{document}